
\magnification=1200
\hsize=5.85truein
\vsize=8.15truein
\hoffset 0.525truein
\voffset 0.25truein
\baselineskip=14pt
\parindent 1.5em
\vskip 1.truecm

\rightline {McGill/94-23}

\bigskip
\bigskip

\centerline{\bf Elements of Finite Order in Lie Groups}

\centerline{\bf and Discrete Gauge Symmetries}

\vskip 1.5truecm

\centerline{M. de Montigny}

\vskip 0.5truecm

\centerline {Department of Physics\/}

\centerline {McGill University\/}

\centerline {Montreal, Quebec\/}

\centerline {H3A 2T8, CANADA\/}

\vskip 1.truecm

\centerline{ABSTRACT}

We apply Kac's theory of elements of finite order
 (EFO) in
 Lie groups to the description of
 discrete gauge symmetries in various
 supersymmetric grand unified models. Taking
 into account the discrete anomaly cancellation conditions,
 we identify the EFO which generate certain matter parities in the
 context of the supersymmetric $SO(10)$ and $E_6$ models.

\bigskip


\bigskip
\bigskip
\bigskip

Submitted to: {\it Nucl. Phys. B}.

\vfill
\eject

\noindent {\bf 1. Introduction.}
\medskip

The study of discrete symmetries has found many applications in
 particle physics. Among their prominent phenomenological issues, let
 us mention that they arise in solar neutrino models, and
 provide a
 constrained structure to the quark and lepton mass matrices,
 suppression of FCNC,
 and elimination of terms in the superpotential
 which could lead to pathological phenomena such as a too rapid
 proton decay.
 However, global discrete symmetries experience drawbacks at the
 Planck scale, because they do not survive some gravitational
 ({\it e.g.} wormhole) effects. (For a review, see Ref. [1]).
 This, in addition to the theorists'
 prejudice in favor of local symmetries,
 provides a
 natural reason to promote them to ``discrete gauge symmetries''
 (DGS).
 According to the Krauss-Wilczek
 scenario, DGS appear as the remnants of a
 given continuous gauge group [2]. (It has been discussed recently that
 CP could be interpreted as a DGS [3].)

 The simplest example consists in the
 spontaneous breakdown of a $U(1)$ gauge symmetry by a Higgs field
 of charge $Nq\neq 0$. If the $U(1)$ charge of every particle of the
 model is an integer multiple of $q$ (not necessarily the
 charge of one of the fields) then a $Z_N$ discrete symmetry
 survives. Similarly, here we consider only discrete
 {\it abelian} groups, although the elements of finite order
 of a Lie group generate
 {\it nonabelian} discrete groups as well.
 It is not always possible to gauge any discrete symmetry
 because, for instance, of the requirement of anomaly cancellation
 [4, 5].
 Constraints have been discussed and classified {\it e.g.} in Refs.
 [6-8].

When the discrete group is a cyclic group $Z_N$, its action on a
 multiplet is made manifest through the assignment of complex
 roots of unity to the
 fields of the multiplet. These complex
 numbers can be seen as the eigenvalues of an automorphism of finite
 order of the gauge group acting on the fields representation.
 Given a continuous gauge group $G$ ({\it e.g.} a
 grand unified group) which
 contains the remnant DGS, the action of the discrete group
 on a multiplet of $G$ is the same as the action of an automorphism
 of an {\it element of finite order} $X\in G$ ({\it i.e.}
 $X^N = Id_G$, for some $N$).
 Obviously, for any automorphism $\phi$ of $X$
 which acts on a multiplet of $G$, we have
 $$
 \phi^N(X)=X,$$
 so that $\phi$ provides eigenvalues to the fields, and thus
 defines a $Z_N$ symmetry. Here, whether the DGS are
 contained or not in a $U(1)$ is immaterial.
 As for the usual case where the $U(1)$
 subgroup of $G$ must be not embedded into the standard model gauge
 group, here the set of elements of finite order must not be
 contained in the standard model.

Here we restrict $\phi$ to be an {\it inner} automorphism of $G$
 ({\it i.e.} $\phi$ has the form $\phi_Y(X)=YXY^{-1}$, for some
 $Y\in G$), and use Kac's formalism of
 elements of finite order (EFO), which has known important
 developments and applications [9, 11] since its original formulation
 in Ref. [10].
 The theory of EFO has provided valuable tools in calculating
 with representations of Lie groups.
 For example, their ``characters'' provide an efficient and
 practical way for determining the irreducible components of the
 tensor product of irreducible representations. Also, they can
 be used to generate signatures of representations of noncompact
 real forms. (For other applications, see Refs. [9-11]).

 For our purpose, the main interest in EFO theory
 is that they provide a systematic way to
 assign eigenvalues of finite order ({\it i.e.} of type
 $exp\left ( {2\pi i\over N}k \right )$, where $k$ is an
 arbitrary integer, and $N$ is a positive integer) to the weight
 vectors of a representation or, equivalently, to the various
 superfields in a multiplet. In the mathematics literature, the
 EFO is said to provide a ``$Z_N$ grading'' of the representation,
 that is, a decomposition of the representation space,
 $$
 V=V_0 + V_1 + \cdots,$$
where the subspace $V_l$ corresponds to the eigenvalue
 $exp\left ( {2\pi i\over N}l\right )$, along with a
 decomposition of the Lie algebra,
$$
 g=g_0 + g_1+\cdots,$$
 with $g_k$ associated to
 $exp\left ( {2\pi i\over N}k\right )$, such that
$$
 g_k\cdot V_l = V_{k+l},$$
where $g_k$ and $V_l$ represent any of their respective elements.
 Therefore we would like to underline the fact that,
 in the context of a
 grand unified model, a discrete symmetry provides a grading of the
 unification {\it Lie algebra} as well
 as the representation. This is
 analogous to the fact that the time reversal operation leaves
 unchanged the angular momenta ${\bf J}$ and space translation operators
 ${\bf P}$,
 but multiply by $-1$ the time translation $H$ and Lorentz boosts
 generators ${\bf K}$, thus providing the Poincar\'e
 algebra with a $Z_2$ grading:
$$
[g_\mu , g_\nu ]=g_{\mu +\nu\ mod 2};\
 g_{0} = \{ {\bf J}, {\bf P}\},\
 g_{1} = \{ H, {\bf K} \}.$$
 Another example consists of the gradings provided by the
 action of ``generalized charge conjugation operators'',
 introduced in Ref. [12].
 Apart from being naturally connected with the existence of finite
 groups in Lie groups, EFO allows one to determine the ``congruence
 class'' of an irreducible representation,
 which has been considered in Ref. [13]
 as a criterion for discrete gauge $R$-parity to survive at low
 energy. (We shall consider this point later in the paper). Thus we
 might expect the theory of EFO to be a useful tool for the
 investigation of discrete symmetries in various contexts of
 particle physics.

The purpose of this paper is to bring attention to Kac's
 theory of EFO for the
 study of DGS in grand unified theories.
 As
 an example, we identify
 some generalized matter parities with EFO.
 This method could
 certainly find applications in other areas of particle physics,
 given its natural relationship with discrete groups.
 In Section 2, we provide an elementary introduction to the
 theory of EFO, with emphasis on the points that we need. The
 method is restricted to the {\it diagonal representative} of a
 conjugacy class of EFO. As an illustration
 of the method, in Section 3 we identify the EFO associated to
 anomaly free DGS which ensure
 proton stability in supersymmetric models.
 We make some concluding
 remarks in Section 4. As we will mention in this paper, the
 method in its present form
 does {\it not} allow one to identify all the
 discrete subgroups of a Lie group,
 although an elegant extension going
 into that direction seems possible. In fact, {\it all} the discrete
 subgroups of a Lie group are generated by EFO, in general
 non-diagonal.
 Also, since we are considering
 subgroups of a Lie group, the present method cannot describe
 discrete symmetries which are not in some
 (grand unification) Lie group.
 Therefore, the discrete groups that we describe must
 be the remnants of some grand unified groups.

\bigskip

\noindent {\bf 2. Overview of EFO theory.}
\medskip

In this section, we recall some basic
 facts about the theory of EFO. (Details can be found in
 Refs. [9-11]).
 An EFO is an
 element of a Lie group $G$ (which we assume to be simply
 connected), and acts on its Lie algebra via the adjoint
 representation.
 An element $X\in G$ is of order $N$ if it is the smallest positive
 integer such that
$$
X^N=Id_G. \eqno(1)$$
This implies, in particular, that the {\it inner}
 automorphisms (like any automorphism) of the Lie algebra
 $g$ of $G$,
$
\phi (x) = XxX^{-1}, x\in g $
 are of finite order ({\it i.e.} $\phi^M(x)=x$, for $M$ some
 divisor of $N$).
 Hence, $\phi$ splits
 $g$ as $g=\oplus_{k=0}^{N-1} g_k$, according to
$$
\phi (x_k) = exp\left ( {2\pi i\over N}k\right ) x_k, \eqno(2)$$
 where $x_k$ lies in the vector subspace $g_k$ of $g$.
 This decomposition is
 a $Z_N$ {\it grading} of $g$
 [9, 10]. The cyclic group $Z_N$ is generated
 by $X$.

For example, the element of $SU(3)$,
$$
X=\pmatrix {-1 & 0 & 0\cr 0 & 1 & 0\cr 0 & 0 & -1\cr},\eqno(3)$$
is of order 2 ({\it i.e.} $X^2=Id_{SU(3)}$),
 and acts on the Lie algebra $su(3)$ as
$$
\eqalign {
X \pmatrix {h_1 & e_{\alpha_1} & e_{{\alpha_1} +{\alpha_2}}\cr
 e_{-{\alpha_1}} & h_2 & e_{\alpha_2}\cr
 e_{-({\alpha_1} +{\alpha_2})} & e_{-{\alpha_2}} & -(h_1+h_2)\cr}
& X^{-1}= \cr
& \pmatrix {h_1 & -e_{\alpha_1} & e_{{\alpha_1} +{\alpha_2}}\cr
 -e_{-{\alpha_1}} & h_2 & -e_{\alpha_2}\cr
 e_{-({\alpha_1} +{\alpha_2})} & -e_{-{\alpha_2}} & -(h_1+h_2)\cr},
 \cr }\eqno(4)$$
so that $su(3)$ is partitioned into two components,
$$
su(3)_0 =
\pmatrix {h_1 & 0 & e_{{\alpha_1} +{\alpha_2}}\cr
 0 & h_2 & 0\cr
 e_{-({\alpha_1} +{\alpha_2} )} & 0 & -(h_1+h_2)\cr},
\ \
su(3)_1 =
\pmatrix {0 & e_{\alpha_1} & 0\cr
 e_{-{\alpha_1}} & 0 & e_{\alpha_2}\cr
 0 & e_{-{\alpha_2}} & 0\cr}. \eqno(5)$$
The elements of component $su(3)_0$ belong to the eigenvalue
 $(-1)^0$ in (4), and those of $su(3)_1$, to
 eigenvalue $(-1)^1$.

We have described the decomposition (or {\it grading})
 of a Lie algebra. We need also to consider
 the action of EFO on representations.
 For a representation $(\rho ,V)$ of $G$, such that
 $v'=\rho (X)\cdot v,\ \ X\in G$ being an EFO,
 and $v, v'\in V$, we have the analog of
 (1),
$$
\rho (X)^{N(\rho )} = Id_{End(V)},\eqno(6)$$
where $N(\rho )$ is some divisor of $N$.
 (From now on, we consider representations of
 Lie {\it algebras}, rather than
 Lie {\it groups}). As for the Lie algebras,
 eigenvalues are thus assigned to representation vectors $v$
 belonging to the representation space $V$ of $g$ as
 follows,
$$
\rho (X)\cdot v_k = exp\left ({2\pi i\over N}k\right )v_k, \eqno(7)$$
where $v_k$ is an element of the representation subspace $V_k$.
In that case, the representation space
 is graded compatibly
 with the grading of $g$ obtained previously. This means that if
 $x$ is in the subspace $g_k$ of $g$, and $v$ is in the subspace
 $V_l$ of $V$, then $v'=\rho (x)\cdot v$ is either zero, or
 in the representation subspace
 $V_{k+l\ (mod N)}$.

For example, the element (3) of $SU(3)$ partitions the
 three-dimensional irreducible representation of $su(3)$
 as
$$
X\pmatrix {a\cr b\cr c\cr} = \pmatrix {-a\cr b\cr -c\cr}
\rightarrow
V_0 = \pmatrix {0\cr b\cr 0\cr},\ V_1 = \pmatrix {a\cr 0\cr c\cr}.
\eqno(8)$$
This $Z_2$ grading is compatible with
 that of $su(3)$ found in (5), that is,
 $$su(3)_0\cdot V_0= V_0,\
 su(3)_0\cdot V_1= V_1,\ su(3)_1\cdot V_0= V_1,\
 su(3)_1\cdot V_1= V_0.$$
In practice, we are much more interested in the decomposition
 (8) of the representation than in the Lie algebra itself (5). It
 is such a grading which achieves the phenomenological virtues
 mentioned in the introduction.

We now turn to the description of Kac's algorithm and notation [9-11].
 Different elements of $G$ which belong to the same conjugacy class have
 the same order. Hereafter the objects of interest are the conjugacy
 classes of EFO, and we shall give a way to obtain their unique
 diagonal representative.
 For every element $X$ of the Lie group $G$ there is
 a unique element $x\in F$ (where $F$ denotes the
 fundamental region of the affine Weyl group acting on the Cartan
 subalgebra of the Lie algebra $g$) such that $X$ is conjugate to
 $exp \left ( 2\pi i x\right )$. If $X$ is an element of finite
 order, we denote by ${\bf s}$ the point in $F$ which represents the
 conjugacy class of $X$.
 Hence, every conjugacy class of EFO meets $F$ in
 exactly one point. According to Kac's notation,
 the point {\bf s\/} is labelled by a set of weighted barycentric
 coordinates
$$
{\bf s\/} = [s_0, \dots ,s_r],$$
where the greatest common divisor of the $s_i$ is 1, and $r$ is the
 rank of $G$. The sets ${\bf s}$ are in one-to-one correspondence with
 the conjugacy classes of EFO in $G$, and the numbers
 $s_0,\dots ,s_r$ are associated with the extended Coxeter-Dynkin
 diagram.
 From now on, the term EFO will actually designate
 a {\it conjugacy class} of EFO, unless stated otherwise.

The useful fact is that every EFO ${\bf s}$
 has within its
 conjugacy class a {\it unique diagonal} representative which acts on
 the representation subspace $V^\lambda$ which belongs to the weight
 $\lambda = \sum_{i=1}^r c_i \alpha_i$ (the $c_i$ being rational numbers,
 and $\alpha_i$, simple roots of $g$) as
$$
v_\lambda \rightarrow
v'_\lambda = exp \left ( {2\pi i\over M}<\lambda ,{\bf s\/}>\right )
 v_\lambda ,
\eqno(9)$$
with $<\lambda ,{\bf s\/}> = \sum_{i=1}^r c_i s_i$. The exact order of
 $X$ is given by $N=MC$, where $M$ and $C$ depend on $g$ and are given in
 Table 6 of Ref. [9]. To summarize, a conjugacy class of EFO
 (denoted by an array ${\bf s}$) provides each weight vector
 ({\it i.e.} particle) in some
 irreducible representation of $G$ with an eigenvalue given by the
 exponential in (9). In other words, equation (9) provides the
 diagonal representative of the class ${\bf s}$.

For example, consider once again
 $SU(3)$. An EFO of $SU(3)$
 is specified by ${\bf s\/} = [s_0 ,s_1 ,s_2]$,
 and its order is $N=MC$, where $$M = s_0 + s_1 +s_2,\ \ \
 C = {3\over {gcd(3;s_1 + 2s_2)}}.$$  The only (conjugacy class of)
 element of order two is given by ${\bf s\/} = [0 ,1 ,1]$, for which
 $M = 2$ and $C = 1$.
 The action of this EFO on the three-dimensional fundamental
 representation of highest weight $(1, 0)$, with the three weight
 vectors $$\lambda_1 = (2\alpha_1 +\alpha_2 )/3,\
 \lambda_2 = (-\alpha_1 + \alpha_2)/3,
 \ \lambda_3 = (-\alpha_1 - 2\alpha_2 )/3,$$
 is given, using (9), by
 $$v_{\lambda_1}\rightarrow -v_{\lambda_1},
 \ v_{\lambda_2}\rightarrow v_{\lambda_2},
 \ v_{\lambda_3}\rightarrow -v_{\lambda_3}.\eqno(10)$$
 Written as a matrix,
 this is nothing but (3). In the same way, the adjoint representation
 ({\it i.e.} $su(3)$ itself) has its basis elements graded as
$$
su(3)_0=\{h_1, h_2, e_{\pm ({\alpha_1} +{\alpha_2} )}\},\ \
su(3)_1=\{e_{\pm{\alpha_1}},e_{\pm{\alpha_2}}\},$$
in accordance to (5).

Below we shall find all EFO of order $2$ and $3$, in supersymmetric
 grand unification models $SO(10)$ and $E_6$.
 We denote each
 EFO by ${\bf s}_{(i)}$ (not to be confused with its {\it components}
 $[s_0, \dots ,s_r]$). Suppose that there are $m$ EFO
 ${\bf s}_{(1)},\dots ,{\bf s}_{(m)}$ of order $N$. They lead
 {\it a priori} to $N^m$ different discrete actions on multiplets.
 An EFO ${\bf s}_{(j)}$ acts on the vector belonging
 to the weight $\lambda$ as (see Refs. [9-11]):
 $$
 v_\lambda \rightarrow v^\prime_\lambda =
 exp\left ({2\pi i\over M}<\lambda , {\bf s}_{(j)}>\right )
 v_\lambda .\eqno(11)$$
 Given $m$ EFO: ${\bf s}_{(1)},\dots ,
 {\bf s}_{(m)}$,
 we shall denote the product of their eigenvalues by
 ${\bf s}_{(1)}^{n_1}\cdots{\bf s}_{(m)}^{n_m}$ (where the
 $n_i$'s are nonnegative integers), and its action is given
 by
$$
\eqalign {v_\lambda\rightarrow v'_\lambda
&=\left (e^{{2\pi i\over M}<\lambda ,{\bf s}_{(1)}>}\right )^{n_1}
 \cdots
\left (e^{{2\pi i\over M}<\lambda ,{\bf s}_{(m)}>}\right )^{n_m}
 v_\lambda ,\cr
 &= exp \left (
{{2\pi i\over M}\left (n_1<\lambda ,{\bf s}_{(1)}> +\cdots
 + n_m<\lambda ,{\bf s}_{(m)}>\right )}\right ) v_\lambda ,\cr
 &= exp \left ({{2\pi i\over M}
 \sum_{k=1}^m n_k <\lambda ,{\bf s}_{(k)}>} \right )
 v_\lambda ,  \cr }\eqno(12)
$$
for all the weights $\lambda$ of the representation.

The {\it product} ${\bf s}_{(1)}^{n_1}\cdots
 {\bf s}_{(m)}^{n_m}$
 allows one to identify
 a discrete symmetry acting on a representation space.
 Let us mention that the symmetries obtained here are just
 those obtained from {\it inner} automorphisms of a Lie
 algebra, from elements of finite order in the corresponding
 Lie group. Hence we do not find all the discrete components
 admitted by a Lagrangian invariant under the
 Lie group. In order to get the complete set of discrete
 symmetries, one should also consider the action of
 {\it outer} automorphisms of the Lie algebra. Such a study
 would require the use of the present algorithm by taking an
 EFO in a group larger than $G$ ({\it i.e.} in which $G$ is
 contained. In practice, we could take $G\times G$, for
 example). As mentioned by Moody-Patera-Sharp [11], any
 discrete group of $G$ consists of elements of finite order,
 so that any element
 of that finite group can be identified with elements
 of finite order in $G$. However the prescription described
 previously only allows to describe
 the diagonal representatives of {\it conjugacy classes}
 of EFO. The problem of identifying
 systematically all the discrete groups in some
 Lie group by means of EFO has not been considered in the
 literature yet. Here we shall not pursue this problem
 further and hence, our list of DGS is not complete.

Next we will
 take into account various constraints in order to obtain pertinent
 discrete symmetries. For instance, if the symmetry is color-blind then
 the three weight vectors corresponding to each quark flavor must have
 the same eigenvalue. Also, one must take into account anomaly
 cancellation constraints [6, 7]. This is discussed
 in the next section.

\bigskip

\leftline {\bf 3. Application to matter parities.}

\medskip

In this section, we interpret the grading described
 in the previous section as being the result of a discrete symmetry.
 Specifically, we consider EFO contained in some unification Lie
 group (in a supersymmetric model) which acts
 on some multiplets.
 We must impose some restrictions upon the possible gradings. For
 instance, if a discrete symmetry is color-blind, that is, if it acts
 in the same way upon the various colors of a fixed quark flavor,
 then we must consider EFO (or {\it products} of EFO) which provide
 the same eigenvalue to the different colors of a given quark.
 Also, since a DGS is a relic of a spontaneously broken continuous
 gauge symmetry, then it should satisfy certain anomaly
 cancellation conditions. That imposes further constraints
 on the possible EFO. We consider all of these constraints
 in the present section.

A $Z_N$ symmetry acts on the fields (or superfields) as
$$
\eqalign
 {
\Psi_k\rightarrow \Psi^\prime_k &= \omega_N^{(k)}\Psi_k,\cr
&=exp\left ({2\pi i\over N}\alpha_k\right )\Psi_k ,\cr}
\eqno(13)$$
where the $\alpha_k$ are some
  charges, which can be identified with the product of
 EFO eigenvalues (12) as
$$
\alpha_k = {N\over M}\sum_{l=1}^m
 n_l<\lambda ,{\bf s}_{(l)}> ,\eqno(14)$$
where the field $\Psi_k$ corresponds to the weight vector
 $\lambda$.
 Like the authors of Ref. [14], we consider the
 three generators $\omega
 = R, A$ and $L$, where
$$
R_N=exp \left ({2\pi i\over N} I_3^R \right ),\
 A_N=exp \left ({2\pi i\over N} Y_A \right ),\
 L_N=exp \left ({2\pi i\over N} L \right ).\eqno(15)$$
In other terms, the charges $\alpha$ in (13) are
 $I_3^R, Y_A$ and $L$.
The charge $I_3^R$ corresponds to the third component of a right-handed
 weak-isospin, $Y_A$ couples the quarks and leptons like the $E_6$
 generator corresponding to the (Cartan) diagonal generator of the
 $SU(2)$ subgroup appearing in the decomposition $E_6\supset
 SU(6)\times SU(2)$, and $L$ is the standard lepton number.
 (Their charges for the standard superfields are listed in Table I.)
 A general element $g_N$
 of a $Z_N$ discrete symmetry can be written:
$$
g_N = R_N^m\times A_N^n\times L_N^p,\ \ \
 m, n, p = 0,\dots ,N-1,\eqno(16)$$
so that $R, A, L$ form a ``basis'' of DGS.

As mentioned in [14], these symmetries are further constrained by
 the imposition of discrete anomaly cancellation conditions [6].
 Necessary (but not sufficient) conditions are given in Ref. [6]
 for:
$(1)$ cubic $Z_N^3$;
$(2)$ mixed $Z_N$-graviton-graviton [15];
 and $(3)$ mixed $Z_N-SU(M)-SU(M)$ anomalies.
 We consider discrete symmetries which
 are remnants of some grand unification model (a good reference
 about grand unified theories is Ref. [16]). For example,
 the ``matter parity'' is added in the minimal
 supersymmetric extension of the standard model (MSSM) in order
 to eliminate both the baryon- and the lepton-number violating
 terms  and lead to an acceptable rate of proton decay [17].
 The concept of matter parity has been extended to that
 of ``generalized parity'' in Refs. [18] (see also Ref. [14]).
 The scheme of Ref. [2] suggests that these DGS are subgroups
 of a $U(1)$. In turn, the $U(1)$ is a subgroup of a grand
 unified Lie group, but not contained in the standard model gauge
 group. Here, we do not need the intermediate $U(1)$.
 The existence of DGS in various grand unified models has been
 investigated in Ref. [19].

In their study of dimension four operators in the superpotential,
 Ib\'a\~nez and Ross [14] have found various types of symmetries,
 out of which
 only two generalized parities are discrete anomaly
 free (with the minimal content of the supersymmetric standard
 model). One is
 $(1)$ the standard $R$-parity (which is a $Z_2$
 symmetry given by $R_2$ in (15)), and the other is
 $(2)$ the $Z_3$ symmetry generated by $B_3=R_3L_3$ [14].
 (There are other anomaly free $Z_3$ symmetries ($R_3, L_3, R_3 L_3^2$)
 but they require the existence of additional fermions with fractional
 $Z_3$ charge, so that the actual symmetry is rather --at least--
 $Z_9$ [8].)

We now study the presence of the discrete symmetries (15)
 generated by EFO in the
 supersymmetric grand unified models $SO(10)$ and $E_6$.
 We identify the conjugacy classes of EFO which provide us with
 some generators of the
 Ib\'a\~nez-Ross [14] classification of discrete symmetries.
 The flavor content of the weight systems of interest for $SO(10)$
 and $E_6$ is in the Tables I, II and V
 of Ref. [19]. The order $N$ of an EFO is given by
 $N=MC$. For $SO(10)$, an EFO
 ${\bf s\/}=[s_0, s_1, \dots ,s_5]$ has
 $$
 M=s_0+s_1+2s_2+2s_3+s_4+s_5, \eqno(17)$$
 and
 $$
 C={4\over {gcd(4;2s_1+2s_3+3s_4+5s_5)}}. \eqno(18)$$
 For $E_6$, an EFO is specified by
 ${\bf s\/}=[s_0, \dots ,s_6]$ and has
 $$
 M=s_0+s_1+2s_2+3s_3+2s_4+s_5+2s_6,\eqno(19)$$
 and
 $$
 C={3\over {gcd(3;s_1-s_2+s_4-s_5)}}. \eqno(20)$$
 $M$ is the order of the action on the adjoint representation,
 {\it i.e.} $N(\rho )$ in (6) for $\rho =$ adjoint representation.
 Values of $M$ and $C$ for other simple Lie groups are
 given in Table 6 of Ref. [10].

The first point we must investigate is to find all the EFO of
 a given order $N$.
 Let us first consider the $Z_2$ discrete symmetries. For $SO(10)$ there are
 three EFO of order two, namely
 $${\bf s}_{(1)}=[0,0,0,0,1,1],\ {\bf s}_{(2)}=[0,0,1,0,0,0],\
 {\bf s}_{(3)}=[0,1,0,0,0,0],\eqno(21)$$
as can be verified by using (17) and (18). In order to find the
 products (12) that lead to the generators $R, A, L$ of (15), one could,
 in principle, solve (14) for $n_l$, but notice that this is not a
 simple algebraic system, because both sides of the equation actually
 read modulo $N$.
 The product ${\bf s}_{(1)}{\bf s}_{(2)}{\bf s}_{(3)}$
 provides the discrete symmetry $R_2$.
 In other words, each ${\bf s}_{(i)}$ corresponds to an
 element of order two of $SO(10)$ (with its action on the weight
 vectors given by (9)), and the product is the actual
 product of the three group elements, represented by diagonal
 matrices.

 For $E_6$, the situation is less interesting. There are two EFO
 of order two:
 $${\bf s}_{(1)}=[0,1,0,0,0,1,0],\ \
{\bf s}_{(2)}=[0,0,0,0,0,0,1].\eqno(22)$$
 Unlike the case of $SO(10)$, they do not reproduce any of the
 generator in (14). At best,
 their product provides the discrete symmetry $R_2$, but after
 multiplying the Higgs fields $h, h'$ by $-1$.

 We now turn to the $Z_3$ discrete symmetries. For $SO(10)$,
 there are six EFO of order three:
 $$\eqalign{ {\bf s}_{(1)}&=[1,2,0,0,0,0],\
\  {\bf s}_{(2)}=[1,0,1,0,0,0], \cr
 {\bf s}_{(3)}&=[0,1,0,1,0,0],\ \  {\bf s}_{(4)}=[0,1,0,0,2,0], \cr
 {\bf s}_{(5)}&=[0,1,0,0,0,2],\ \  {\bf s}_{(6)}=[1,0,0,0,1,1]. \cr }
 \eqno(23)$$
 The discrete symmetry $\alpha_A$ is given by the products:
 $${\bf s}_{(2)}{\bf s}_{(3)}^2{\bf s}_{(4)}^2
{\bf s}_{(5)}{\bf s}_{(6)}^2,\
 {\bf s}_{(1)}{\bf s}_{(2)}{\bf s}_{(3)}^2{\bf s}_{(4)}
{\bf s}_{(6)},\ {\bf s}_{(1)}{\bf s}_{(2)}^2{\bf s}_{(3)}{\bf s}_{(5)}.
$$
The symmetry $\alpha_R$ corresponds to any of
 the two possibilities:
 $$
 {\bf s}_{(1)}{\bf s}_{(2)}{\bf s}_{(3)}{\bf s}_{(6)},\
 {\bf s}_{(1)}
{\bf s}_{(2)}^2{\bf s}_{(3)}^2{\bf s}_{(4)}{\bf s}_{(5)}.
 $$

 For $E_6$,
 there are five EFO of order three:
 $$\eqalign{ {\bf s}_{(1)}&= [0,0,0,1,0,0,0],\ \
 {\bf s}_{(2)}=[1,0,0,0,0,0,1], \cr {\bf s}_{(3)}&=[1,1,0,0,0,1,0],\ \
 {\bf s}_{(4)}=[0,1,1,0,0,0,0], \cr {\bf s}_{(5)}&=[0,0,0,0,1,1,0].\cr }
 \eqno(24)$$
 We have found that
 ${\bf s}_{(1)}{\bf s}_{(2)}{\bf s}_{(3)}^2
{\bf s}_{(4)}{\bf s}_{(5)} $
 provides exactly the discrete symmetry $\alpha_R$. The discrete
symmetry $\alpha_A$ is given by the product
 ${\bf s}_{(1)}{\bf s}_{(2)}^2{\bf s}_{(3)}{\bf s}_{(4)}^2
{\bf s}_{(5)}$. (Obviously, we can form combinations like
 $\alpha_A\alpha_R={\bf s}_{(1)}{\bf s}_{(5)}$, using the
 property ${\bf s}_{(k)}^3=Id$.)

In all cases, the generator $L$ does not correspond
 to any EFO. As explained before, Kac's notation provide the
 diagonal representative of a conjugacy class of EFO, so that
 the missing DGS could be generated by non-diagonal elements
 of finite order. This is not further studied here.
 This restriction of the method prevents us from finding
 the symmetries which forbid FCNC, since the most general
 form of such symmetries is
$$
S_N=R_N^m\times\left (L_N^j\right )^{p_j}\times A_N,\eqno(25)$$
where $L^j$ corresponds to the three standard lepton numbers. In
 principle, one could need just to consider three copies of the
 fermions representations, with the same element of finite order
 generating $L_N^j$. (Generation dependent DGS are discussed in
 Ref. [20].) Hence, of the two anomaly free generalized parities
 $R_2$ and $R_3L_3$ [14], only $R_2$ can be found by Kac's method
 in its current form, in the $SO(10)$ model.

To close this section, let us recall from Ref. [10] the relation
 between congruence classes of representations and the theory
 of EFO. Having in mind the criterion stated in Ref. [13], based
 on the congruence classes of representations, let us say that in
 the present framework, they can be obtained from the EFO associated
 to a central element of the Lie group $G$. In turn, a central
 element is given by an EFO whose entry is at a ``tip'' of the
 extended Coxeter-Dynkin diagram, that is, a node which has numerical
 mark equal to $1$. For $SO(10)$, the tips are at $s_0, s_1, s_4$,
 and $s_5$. For $E_6$ they are $s_0, s_1$ and $s_5$. Indeed, the center
 of $SO(10)$ contains four elements and there are four congruence
 classes. The center of $E_6$ consists of three elements, and they
 describe the three congruence classes. Therefore,
 we expect Martin's ``surprisingly mild conditions'' [13] to appear as
 natural consequences when DGS are considered in the formalism of
 EFO.

\bigskip

\noindent {\bf 4. Concluding remarks.}
\medskip

Our purpose in doing this work is to bring attention
 to Kac's theory of EFO as a
 mathematical concept to understand formally and generate
 elegantly discrete symmetries in a Lie group. We have applied the
 concept to the DGS in some grand unified models, although the
 procedure could be applied to a much larger class of problems
 in particle physics, where discrete groups are often encountered.

We have found that one cannot generate all the anomaly free
 discrete symmetries, such as they appear in Ref. [14].
 This is so because the method
 describes only the action of the diagonal representatives of
 conjugacy classes of EFO rather than individual
 elements of finite order, which could lead to all (abelian and
 nonabelian) discrete subgroups of a Lie group. Therefore, a possible
 avenue of research is to extend Kac's EFO theory so that
 one could describe as well non-diagonal elements of a conjugacy
 class. Another possible extension would be to treat systematically
 the action of EFO through {\it outer} automorphisms in addition to the
 {\it inner} automorphisms.
 Thus, although the present method encompasses
 many DGS which are not anomaly free, it does not allow to
 find the complete set of anomaly free DGS.
 If such extensions could be achieved, then the present concepts might
 display some elegant properties when applied to particle physics.
 It might be possible that we get such nice properties also by changing
 the assignment of particles to vectors in the representation (so that,
 for instance, one could predict the EFO which give the various parities
 discussed in Section 3.).

An interesting question, related to the generation of discrete
 groups by EFO, is if the EFO theory provides an elegant algorithm to
 tell whether or not a given EFO belongs to some subgroup. For
 instance, in the context of some grand unified theory, it is
 relevant to know whether the discrete group generated by an EFO
 belongs to the standard model subgroup. Put the inverse way,
 given a subgroup ({\it e.g.} the standard model) it would be
 interesting to describe, through EFO, the complete set of
 discrete symmetries which are complementary to this subgroup
 in a larger Lie group ({\it e.g.} grand unified gauge group).
 There is an indication that the presence of zeros in the array
 ${\bf s}$ can be identified with the subgroup associated to the
 corresponding nodes in the Coxeter-Dynkin diagram. This question
 has not been answered in general [J. Patera, {\it private
 communication}].

Another limitation of the present procedure is that it concerns
 only the discrete symmetries which are contained in some Lie
 group. In other words, one cannot get the discrete symmetries
 of some Lagrangian which are not in the continuous group of
 symmetry of the Lagrangian.
 In principle, the method can be applied to other
 grand unified models, like $SU(5), SU(4)\times SU(2)^2,
 SU(3)^3$, etc. or as large as $E_8$ or $SU(16)$.

Other applications in high energy physics have not been explored yet.
\bigskip

\leftline {\bf Acknowledgement\/}
\medskip

I am indebted to Prof. Ji\u ri Patera for introducing to me the
 EFO theory, and
 to Dr Manel Masip for useful discussions about
 discrete gauge symmetries. Drs. Patrick
 Labelle and Robert Foot
 are acknowledged for reading the manuscript.
 This work has been supported by the Natural Sciences and
 Engineering Research Council of Canada.

\bigskip

\leftline {\bf References \/}
\medskip

\noindent [1] T. Banks, {\it Physicalia Magazine} {\bf 12} (1990) 19

\noindent [2] L. M. Krauss and F. Wilczek, {\it Phys. Rev. Lett.\/}
 {\bf 62\/} (1989) 1221

\noindent [3] K. W. Choi, D. Kaplan, and A. Nelson, {\it Nucl. Phys.}
 {\bf B 391} (1993) 515; M. Dine, R. Leigh, and D. MacIntire,
 {\it Phys. Rev. Lett.} {\bf 69} (1992) 2030

\noindent [4] J. Preskill, {\it Ann. Phys.} {\bf 210} (1991) 323

\noindent [5] S. Adler, {\it Phys. Rev.} {\bf 177} (1969) 2426;
 J. S. Bell and R. Jackiw, {\it Nuov. Cim.} {\bf A 60}
 (1969) 47

\noindent [6] L. E. Ib\'a\~nez and G. G. Ross,
 {\it Phys. Lett.\/} {\bf B 260\/} (1991) 291

\noindent [7] J. Preskill, S. P. Trivedi, F. Wilczek, and M. B. Wise,
 {\it Nucl. Phys.\/} {\bf B 363\/} (1991) 207;
 T. Banks and M. Dine, {\it Phys. Rev.} {\bf D 45}
 (1992) 1424

\noindent [8] L. E. Ib\'a\~nez, {\it Nucl. Phys.} {\bf B 398} (1993) 398

\noindent [9] R. V. Moody and J. Patera, {\it SIAM J. Alg.
 Disc. Meth.\/} {\bf 5\/} (1984) 359

\noindent [10] V. G. Kac, {\it Func. Anal. Appl.\/}
 {\bf 3\/} (1969) 252

\noindent [11] R. V. Moody, J. Patera and R. T. Sharp, {\it J. Math.
 Phys.\/} {\bf 24\/} (1983) 2387;
  R. V. Moody and J. Patera, in {\it Proc. 13th Coll.
  Group Theor. Meth. Phys.\/}, ed W. W. Zachary,
 College Park, Maryland (1985) p. 308

\noindent [12] R. V. Moody and J. Patera, {\it J. Math. Phys.}
 {\bf 25} (1984) 2838

\noindent [13] S. P. Martin, {\it Phys. Rev.\/} {\bf D 46\/},
  (1992) R2769, and ``Automatic Gauge {\it R} Parity'', in
 {\it Proc. Int. Workshop Supersymmetry and Unification of
 Fundamental Forces}, Northeastern Univ., Boston (1993);
 hep-ph/9307276

\noindent [14] L. E. Ib\'a\~nez and G. G. Ross,
 {\it Nucl. Phys.\/} {\bf B 368\/} (1992) 3

\noindent [15] R. Delbourgo and A. Salam, {\it Phys. Lett.}
 {\bf B 40} (1972) 381; T. Eguchi and P. Freund, {\it Phys. Rev.
 Lett.} {\bf 37} (1976) 1251; L. Alvarez-Gaum\'e and E. Witten,
 {\it Nucl. Phys.} {\bf B 234} (1983) 269

\noindent [16] G. G. Ross, {\it Grand Unified Theories\/}, (Benjamin-Cummings,
 Menlo Park, 1984)

\noindent [17] G. Farrar and P. Fayet, {\it Phys. Lett.\/}
 {\bf B 76\/} (1978) 575; S. Dimopoulos and H. Georgi,
 {\it Nucl. Phys.\/} {\bf B 193\/} (1981) 150; S. Weinberg,
 {\it Phys. Rev.\/} {\bf D 26\/} (1982) 287; N. Sakai and
 T. Yanagida, {\it Nucl. Phys.\/} {\bf B 197\/} (1982) 83

\noindent [18] L. Hall and M. Suzuki, {\it Nucl. Phys.}
 {\bf B 231} (1984) 219; M. Bento, L. Hall, and G. G. Ross,
 {\it Nucl. Phys.} {\bf B 292} (1987) 400

\noindent [19] M. de Montigny and M. Masip,
 {\it Phys. Rev.} {\bf D 49} (1994) 3734

\noindent [20] D. Kapetanakis, P. Mayr, and H. P. Nilles,
 {\it Phys. Lett.} {\bf B 282} (1992) 95

\vfil\eject

\leftskip=1.8pc
\rightskip=1.pc
\baselineskip=24truept

\vfil\eject

\noindent {\bf Table I.}
$Z_N$  charges of generators $R, A, L$.
$$
\baselineskip=10truept
\vbox{
\tabskip=.6truecm
\halign{\hfil#\hfil&\hfil#\hfil&\hfil#\hfil&\hfil#\hfil
&\hfil#\hfil &\hfil#\hfil &\hfil#\hfil &\hfil#\hfil &\hfil#\hfil
&\hfil#\hfil &\hfil#\hfil \cr
\noalign{\hrule}
\omit&\omit&\omit&\omit&\omit&\omit&\omit&\omit&\omit&\omit&\omit\cr
\omit&\omit&\omit&\omit&\omit&\omit&\omit&\omit&\omit&\omit&\omit\cr
&$q$&$u^c$&$d^c$&$l$&$e^c$&$h$&$h'$\cr
\omit&\omit&\omit&\omit&\omit&\omit&\omit&\omit&\omit&\omit&\omit\cr
\omit&\omit&\omit&\omit&\omit&\omit&\omit&\omit&\omit&\omit&\omit\cr
\noalign{\hrule}
\omit&\omit&\omit&\omit&\omit&\omit&\omit&\omit&\omit&\omit&\omit\cr
\omit&\omit&\omit&\omit&\omit&\omit&\omit&\omit&\omit&\omit&\omit\cr
$\alpha_R$&$0$&$-1$&$1$&$0$&$1$&$1$&
$-1$\cr
\omit&\omit&\omit&\omit&\omit&\omit&\omit&\omit&\omit&\omit&\omit\cr
\omit&\omit&\omit&\omit&\omit&\omit&\omit&\omit&\omit&\omit&\omit\cr
$\alpha_A$&$0$&$0$&$-1$&$-1$&$0$&$0$&$1$\cr
\omit&\omit&\omit&\omit&\omit&\omit&\omit&\omit&\omit&\omit&\omit\cr
\omit&\omit&\omit&\omit&\omit&\omit&\omit&\omit&\omit&\omit&\omit\cr
$\alpha_L$&$0$&$0$&$0$&$-1$&$1$&$0$&
$0$\cr
\omit&\omit&\omit&\omit&\omit&\omit&\omit&\omit&\omit&\omit&\omit\cr
\omit&\omit&\omit&\omit&\omit&\omit&\omit&\omit&\omit&\omit&\omit\cr
\noalign{\hrule}
}}$$
\vskip 2truecm

\vfill\eject\end